\documentstyle[sprocl,epsf]{article}

\newcommand{\PLB}[3]{Phys.\ Lett.\ {B{#1}}, {#2} ({#3})}
\newcommand{\NPB}[3]{Nucl.\ Phys.\ {B{#1}}, {#2} ({#3})}
\newcommand{\PRD}[3]{Phys.\ Rev.\ {D{#1}}, {#2} ({#3})}
\newcommand{\PRL}[3]{Phys.\ Rev.\ Lett.\ {{#1}}, {#2} ({#3})}
\newcommand{\ZPC}[3]{Z.\ Phys.\ {C{#1}}, {#2} ({#3})}

\begin{document}

\thispagestyle{empty}

\begin{flushright}
        CERN-TH/97-355\\
        hep-ph/9712298\\ 
        December 1997\\
\end{flushright}
\vskip 0.8cm

\title{\large\bf  QUARKONIUM POLARIZATION AS A TEST OF
NON-RELATIVISTIC EFFECTIVE THEORY}

\vspace*{1cm}

\author{ M. BENEKE }
\address{ Theory Division, CERN, CH-1211 Geneva 23 }
\vspace{1cm}

\maketitle\abstracts{
I compare current approaches to quarkonium production 
with regard  
to what they tell us about quarkonium polarization. Predictions for 
$J/\psi$ polarization in hadron-hadron and photon-hadron collisions 
are summarized.}

\vspace*{2cm}

\begin{center}
{\em To appear in the Proceedings of the \\
     International Europhysics Conference\\
     on High Energy Physics,\\
     Jerusalem, Israel, 19-26 August 1997}
\end{center}

\newpage

\setcounter{page}{1}


The production of charmonium involves physics at short and long distances. 
The short-distance part is given, uncontroversially, by the production 
cross section for a charm-anticharm-quark pair at small relative velocity 
$v$ of the two quarks. The process through which the $c\bar{c}$ pair binds 
into a particular charmonium state is sensitive to long times 
$\tau\sim 1/(m_c v^2)\sim 1/(500\,\mbox{MeV})$ and therefore it is 
non-perturbative. Various descriptions, with more or less contact to 
QCD, and relying on quite different physical pictures of this process, 
have been proposed, and used, over the years.
The {\em colour-singlet model (CSM)} \cite{SCH94} assumes that 
only those $c\bar{c}$ pairs form $J/\psi$ which are produced in a 
colour-singlet ${}^3\!S_1$ state already at short distances. The 
long-distance part is Coulomb-binding, accountable for by the wave-function 
at the origin. No gluons with energy less than 
${\cal O}(m_c)$ in the $J/\psi$ rest frame are emitted.
The {\em colour-evaporation model (CEM)} \cite{evaporation1} assumes 
that soft gluon emission from the $c\bar{c}$ pair is unsuppressed. The 
colour and spin quantum numbers of the $c\bar{c}$ pair at short distances 
are irrelevant. The long-distance physics is supposed to be described 
by a phenomenological parameter $f_{J/\psi}$, the fraction of `open' 
$c\bar{c}$ pairs below threshold that bind into $J/\psi$. 
The {\em non-relativistic QCD (NRQCD) approach} \cite{NRQCD} synthesizes 
elements of both approaches. $J/\psi$ can be produced from 
$c\bar{c}$ pairs in any colour or angular momentum state at short distances 
but with probabilities that follow definite scaling rules \cite{L92} in 
$v^2$. Soft gluon emission does take place, but 
the interaction of soft gluons with the heavy quarks is determined by 
the NRQCD effective Lagrangian. Spin symmetry holds to leading order 
in $v^2$. There is a price to pay for the more detailed description 
of the long-distance part in NRQCD: 
It depends on (at least) four (rather than one)  
non-perturbative parameters, which have to be extracted from 
experiment. They are $\langle {\cal O}^{J/\psi}_1 ({}^3 \!S_1) \rangle$,  
$\langle {\cal O}^{J/\psi}_8 ({}^3 \!S_8) \rangle$,  
$\langle {\cal O}^{J/\psi}_8 ({}^1 \!S_0) \rangle$,
$\langle {\cal O}^{J/\psi}_8 ({}^3 \!P_0) \rangle$, where the colour 
and angular momentum state indicated refers to the $c\bar{c}$ pair at 
short distances. The precise definition of these matrix elements is 
given in \cite{NRQCD}.

The following deals exclusively with polarization phenomena in 
$J/\psi$ production. We discuss 
predictions for $J/\psi$ production in hadron-hadron and photon-proton 
collisions, based on the CSM and the NRQCD approach. The prediction by 
the CEM is straightforward and universal: Because the model assumes that 
transitions ${}^3\!S_1\leftrightarrow {}^1\!S_0$ are unsuppressed, we 
expect that $J/\psi$ is always produced unpolarized. A polarization 
measurement has various discriminative powers. One can learn 
to what degree spin-flip transitions are suppressed and thereby check the 
basic assumption that distinguishes the CEM from the NRQCD approach. 
Since the octet production matrix elements of NRQCD (see above) lead 
to a polarization pattern different from the CSM, one can learn about 
the importance of colour-octet production mechanisms. In particular,  
production through a ${}^1\!S_0^{(8)}$ state yields unpolarized 
quarkonium. See \cite{BR96,BR2,polarization} for the other 
production channels.\\ 

\noindent 
{\bf $J/\psi$ production in fixed target hadron-hadron collisions.}
Polarization measurements exist for $\psi$ and $\psi^\prime$ production 
in pion scattering fixed target experiments \cite{HEI91}. 
Both experiments observe an essentially flat angular distribution in 
the decay $\psi\to \mu^+ \mu^-$ ($\psi= J/\psi,\psi'$), 
\begin{displaymath}
\frac{d\sigma}{d\cos\theta }\propto 1+ \lambda \cos^2 \theta,
\end{displaymath}
\noindent where the angle $\theta$ is defined as the angle between 
the three-momentum vector of the positively charged muon and 
the beam axis in the rest frame of the quarkonium. The observed values 
for $\lambda$ are $0.02\pm 0.14$ for $\psi'$, measured at 
$\sqrt{s}=21.8\,$GeV in the region $x_F>0.25$ and 
$0.028\pm 0.04$ for $J/\psi$ measured at $\sqrt{s}=15.3\,$GeV 
in the region $x_F>0$.

\begin{figure}[p]
   \vspace{-1.4cm}
   \epsfysize=10cm
   \epsfxsize=7cm
   \centerline{\epsffile{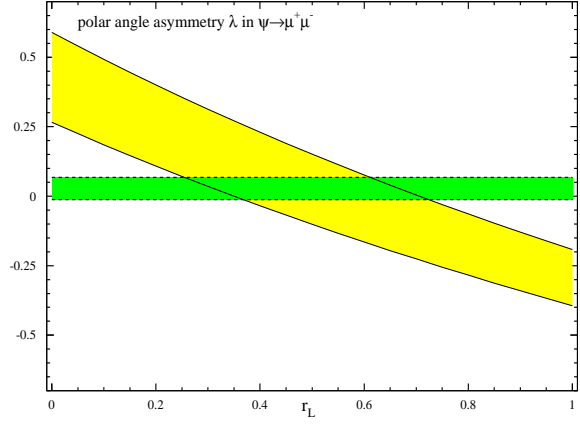}}
   \vspace*{-2.2cm}
\caption[dummy]{Polar angle asymmetry for $J/\psi$ 
production in pion-nucleus collisions at $\sqrt{s}=15\,$GeV 
as a function of the longitudinal polarization 
fraction of indirect $J/\psi$ from radiative feed-down. The horizontal 
band shows the measurement of $\lambda$. \label{ft}}
\end{figure}

The colour-singlet contribution alone yields $\lambda\approx 0.25$ for 
the direct $S$-wave production cross section \cite{FTpred1}. However, the 
total cross section is largely due to colour-octet production. The 
polarization in the colour-octet channels has been considered in 
\cite{BR2} (see also \cite{TAN95}). If 
$\langle {\cal O}_8^{\psi'} ({}^1\!S_0)\rangle$ and 
$\langle {\cal O}_8^{\psi'} ({}^3\!P_0)\rangle$ are 
constrained to be positive, 
$0.15 < \lambda < 0.44$ 
is obtained for $\psi'$ production at $\sqrt{s}=21.8\,$GeV. The lower 
bound is obtained if production through a $c\bar{c}[{}^1\!S_0^{(8)}]$ 
intermediate state dominates. The analysis of $J/\psi$ polarization 
is complicated by indirect $J/\psi$ production through $\chi_c$ 
decays, which are not separated in the measurement above. In 
Fig.~\ref{ft} the polar angle parameter $\lambda$ is plotted as 
a function of $r_L$, the longitudinal polarization fraction of 
indirectly produced $J/\psi$ (i.e. $r_L=1/3$, if $\chi_c$ feed-down gives 
unpolarized $J/\psi$). $r_L$ is difficult to obtain theoretically 
as $\chi_{c1}$ and 
perhaps even $\chi_{c2}$ is dominantly produced through colour-octet 
states, whose polarization yield is described by too many 
phenomenological parameters to be predictable. The wide 
band in Fig.~\ref{ft} is obtained by saturating the direct $J/\psi$ 
production cross section by either 
$\langle {\cal O}_8^{J/\psi} ({}^1\!S_0)\rangle$ (lower curve) or 
$\langle {\cal O}_8^{J/\psi} ({}^3\!P_0)\rangle$ (upper curve). If the 
indirectly produced $J/\psi$ are unpolarized, one would again have to 
assume that 
$\langle {\cal O}_8^{J/\psi} ({}^1\!S_0)\rangle \gg 
\langle {\cal O}_8^{J/\psi} ({}^3\!P_0)\rangle$ in order to reproduce 
the data (horizontal band in Fig.~\ref{ft}). 
A measurement of $r_L$ could clarify the situation.

Since the total cross section is dominated by $J/\psi$ production at small 
transverse momentum, non-factorizable final state interactions may be 
significant (though formally suppressed) and invalidate the predictions 
based on the CSM or NRQCD.\\

\begin{figure}[p]
   \vspace{-1.4cm}
   \epsfysize=10cm
   \epsfxsize=7cm
   \centerline{\epsffile{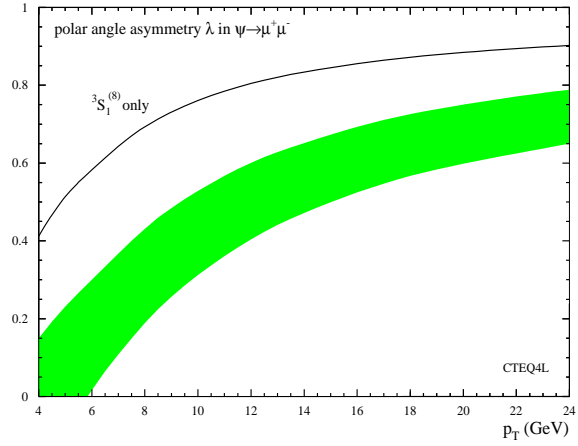}}
   \vspace*{-2.2cm}
\caption[dummy]{$\lambda$ as a function of $p_t$ in 
$p+\bar{p}\to J/\psi+X$ at the Tevatron cms energy $\sqrt{s}=1.8\,$
TeV. From \cite{BEN97}. \label{alpha}}
\end{figure}

\noindent
{\bf $J/\psi$ polarization at the Tevatron.}
At transverse momentum $p_t\gg 2 m_c$ $J/\psi$ production in 
hadron-hadron collisions is now regarded 
as a gluon fragmentation process $g\to c\bar{c}[{}^3\!S_1^{(8)}] \to 
J/\psi+X$ \cite{BF95}. Since the fragmenting gluon is nearly on-shell, 
this implies transversely polarized $J/\psi$ as $p_t\to \infty$ 
\cite{CDFpred} up to spin-symmetry breaking corrections of order 
$v^4$ \cite{BR96}. At finite $p_t$ longitudinally 
polarized $J/\psi$ can be produced, if a hard gluon is radiated 
in the fragmentation process 
\cite{BR96} or the fragmentation approximation is relaxed 
\cite{BEN97,AKL97}. The non-fragmentation terms turn out to be 
particularly important. The predicted polar angle asymmetry 
$\lambda$ is shown in Fig.~\ref{alpha}. At $p_t\sim 5\,$GeV no 
trace of transverse polarization remains. As $p_t$ increases, the 
angular distribution becomes rapidly more anisotropic. The observation 
of this pattern, even qualitatively, would already constitute strong 
support for the gluon fragmentation mechanism {\em and} the 
relevance of spin symmetry in quarkonium production. The polarization 
measurement will therefore 
rule out either the CEM or the applicability of 
NRQCD velocity power counting at the charmonium scale. Because of this, 
this measurement is probably the single most important one that can be 
done in the near future.\\

\begin{figure}[p]
   \vspace{-1.4cm}
   \epsfysize=10cm
   \epsfxsize=7cm
   \centerline{\epsffile{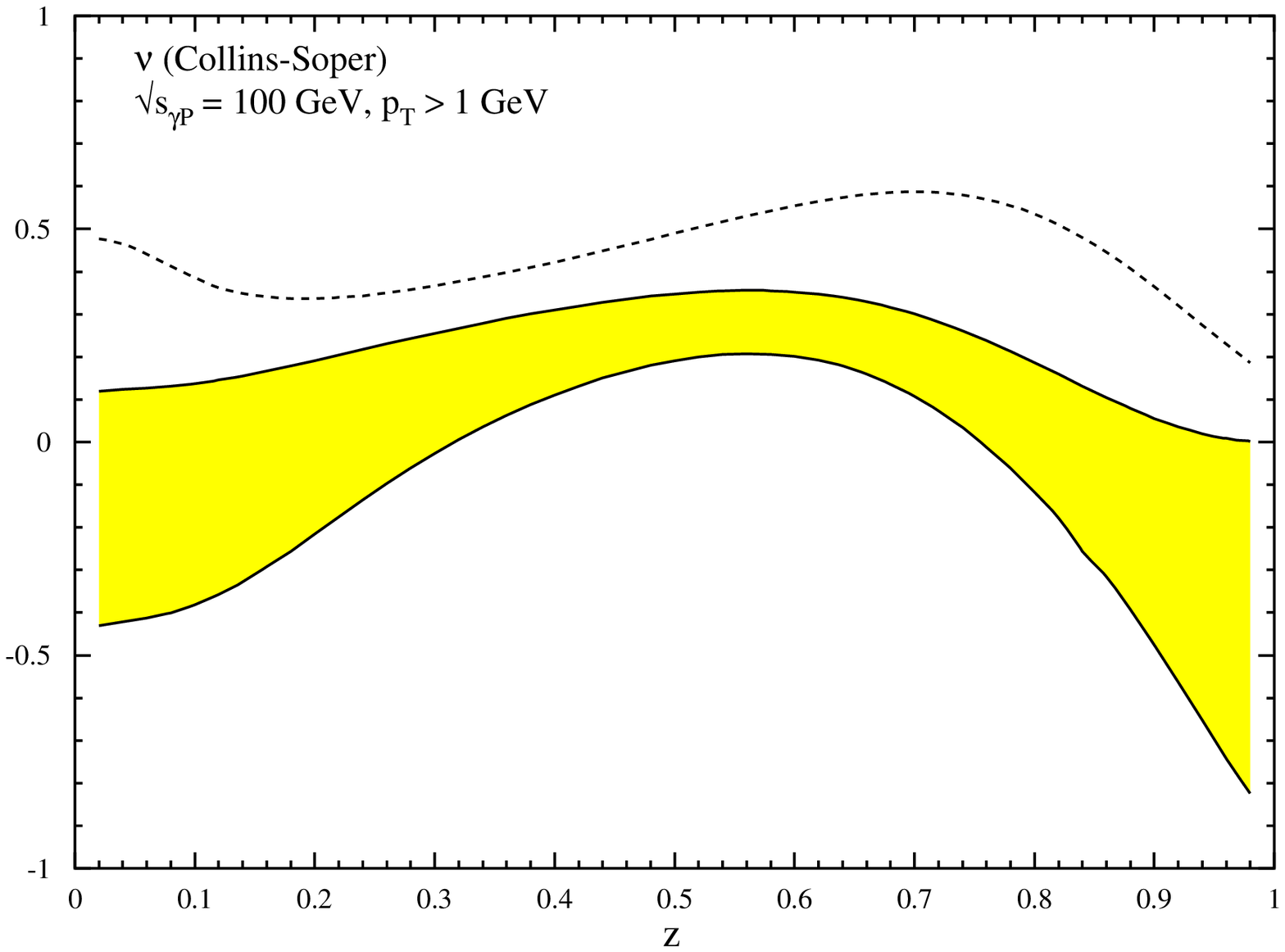}}
   \vspace*{-3.5cm}
   \epsfysize=10cm
   \epsfxsize=7cm
   \centerline{\epsffile{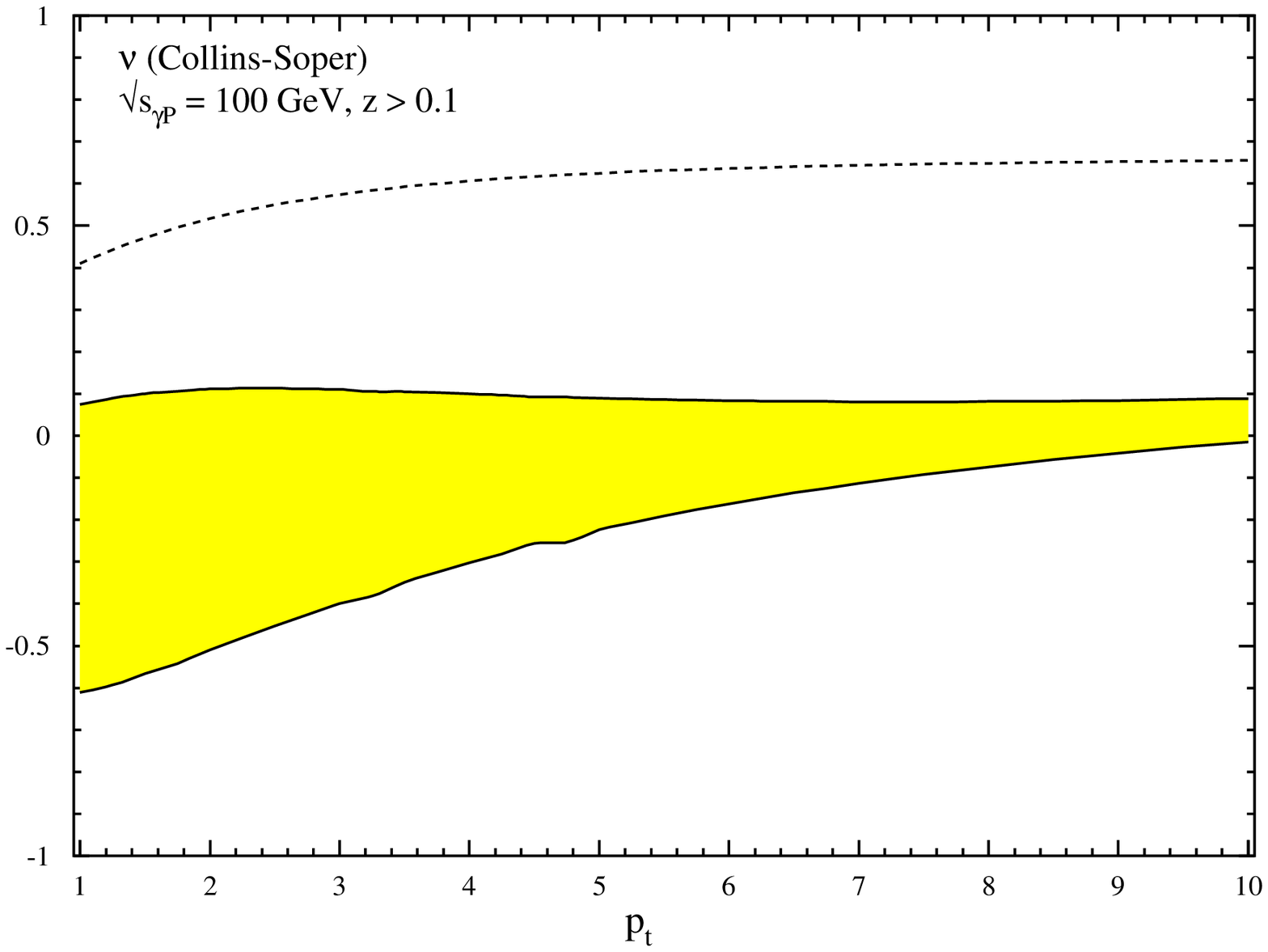}}
   \vspace*{-2.2cm}
\caption[dummy]{Azimuthal angle parameter $\nu$ in the 
Col\-lins\--So\-per frame for $J/\psi$ energy and transverse momentum 
distributions in $J/\psi$ photo\--pro\-duction at a typical HERA energy. 
From \cite{BKV97} where more details can be found. \label{nu_z}}
\end{figure}

\noindent
{\bf $J/\psi$ photo-production at HERA.}
This production process deserves special attention in the context of 
NRQCD (and the CEM), since the colour-octet contributions to the 
energy distribution of inelastically produced 
$J/\psi$ seem to be too large close to the 
point $z=1$ of maximal energy transfer \cite{Cacciari}. Since the 
NRQCD velocity expansion is not valid in this endpoint region 
\cite{BRW}, one would like to infer the relevance of the colour-octet 
contribution from the data themselves. Polar and azimuthal $J/\psi$ decay 
angular distributions may provide a clue to the answer to this problem, 
since the distributions are predicted to be quite different in the CSM 
and in the NRQCD approach \cite{BKV97}. [Recall that all angular 
distributions are isotropic in the $J/\psi$ rest frame in the CEM.] 
The azimuthal dependence, characterized by two additional angular 
parameters $\mu$ and $\nu$, is particularly instructive, as a 
function of both energy fraction $z$ or transverse momentum as 
shown in Fig.~\ref{nu_z}. The shaded band reflects the variation that 
follows if either one of  
$\langle {\cal O}_8^{J/\psi} ({}^1\!S_0)\rangle$,  
$\langle {\cal O}_8^{J/\psi} ({}^3\!P_0)\rangle$ is set to zero, while 
the other saturates the sum which is constrained by the total  
production rate (in other production processes, primarily hadron-hadron 
collisions as discussed above). A measurement of angular distributions 
in inelastic $J/\psi$ production comparable to the measurement of 
polarization in  diffractive $J/\psi$ production at HERA \cite{H1} 
could resolve the controversy whether the measured energy distribution 
is described by the CSM alone and/or is in conflict with the 
size of colour-octet contributions suggested by the NRQCD approach.\\

In the framework of NRQCD, predictions of $J/\psi$ polarization 
have also been obtained for $B\to J/\psi+X$ \cite{FHMN}, 
direct $J/\psi$ production in $Z^0$ decay \cite{LEP} and lepto-production 
of $J/\psi$ \cite{FLE97}. Because of lack of space, the reader is 
referred to the original papers.\\

\noindent {\em Acknowledgements}. I would like to thank M.~Kr\"amer, 
I.Z.~Rothstein and M.~V\"anttinen for their collaboration on this 
topic.

\end{document}